\documentclass[reprint, amsmath,amssymb, aps, prl, a4paper,showkeys, nofootinbib]{revtex4-1}

\usepackage{graphicx} 
\usepackage{amsfonts} 
\usepackage{amsmath} 
\usepackage{amssymb} 
\usepackage{natbib}
\usepackage{dcolumn}
\usepackage{color}
\usepackage{epsfig}

\begin{document}

\title{DebtRank-transparency: Controlling systemic risk in financial networks}

\author{Stefan Thurner$^{1,2,3}$ 
 and Sebastian Poledna$^1$}
\affiliation{
$^1$Section for Science of Complex Systems; Medical University of Vienna; Spitalgasse 23; A-1090; Austria\\
$^2$Santa Fe Institute; 1399 Hyde Park Road; Santa Fe; NM 87501; USA\\
$^3$IIASA, Schlossplatz 1, A-2361 Laxenburg; Austria}

\begin{abstract}
	Banks in the interbank network can not assess the true risks associated with lending 
	to other banks in the network, unless they have full information on the riskiness of 
	{\em all}  the other banks. These risks can be estimated by using network metrics  
	(for example DebtRank) of the interbank liability network which is available to Central Banks. 
	With a simple agent based model we show that by increasing transparency by  
	making the DebtRank of individual nodes (banks) visible to all nodes, 
	and by imposing a simple incentive scheme, that reduces interbank borrowing from 
	systemically risky nodes, the systemic risk in the financial network can be drastically reduced. 
	This incentive scheme is an effective regulation mechanism, that does not reduce the efficiency 
	of the financial network, but fosters a more homogeneous distribution of risk within the system 
	in a self-organized critical way. 
	We show that the reduction of systemic risk is to a large extent due to the massive reduction of 
	cascading failures in the transparent system. An implementation of this minimal regulation scheme 
	in real financial networks should be feasible from a technical point of view. 
\end{abstract}


\maketitle
 
Since the beginning of banking the possibility of a lender to assess the riskiness of a potential 
borrower has been essential. In a rational world, the result of this assessment determines the terms of a 
lender-borrower relationship (risk-premium), including the possibility that no deal would be established in 
case the borrower appears to be too risky. 
When a potential borrower is a node in a lending-borrowing 
network, the node's riskiness (or creditworthiness) not only depends on its financial conditions, but also on those 
who have lending-borrowing relations with that node. The riskiness of these neighboring nodes depends on the 
conditions of their neighbors, and so on. In this way the concept of risk loses its local character between  a 
borrower and a lender, and becomes {\em systemic}.

The assessment of the riskiness of a node turns into an assessment of the entire financial network \cite{haldane_nature}.  
Such an exercise can only carried out with information on the asset-liablilty network. 
This information is, up to now, not available to individual nodes in that network. 
In this sense, financial networks -- the interbank market in particular -- are 
opaque.
This intransparency makes it impossible for individual banks to make rational decisions on 
lending terms in a financial network, which leads to a fundamental principle: 
Opacity in financial networks rules out the possibility of rational risk assessment,  and consequently, 
transparency, i.e. access to system-wide information is a necessary condition for any systemic risk management.   

The banking network is a fundamental building block in our globalized society. It provides a 
substantial part of the funding and liquidity for the real economy \cite{schweitzer_sci}. The real economy -- the ongoing process of invention, 
production, distribution, use, and disposal of goods and services -- is inherently risky. 
This risk originates in the uncertainty of payoffs from investments in business ideas, which might not be profitable,
or simply fail. This type of risk can not be eliminated from an evolving economic system, however it can be spread, shared, and diversified. 
One of the roles of the financial system is to distribute the risk generated by the real economy among the actors in the financial 
network. The financial network can be seen as a service to share the burden of economic risk. 
By no means should this service by itself produce additional systemic risk endogenously.
Neither should  the design and regulation of financial networks introduce mechanisms that leverage or inflate the intrinsic 
risk of the real economy. As long as systemic risk is endogenously generated within the financial network, this system is 
not yet properly designed and regulated. 
In this paper we show, that unless a certain level of transparency is introduced in financial networks,  
systemic risk will be endogenously generated within the financial network. This systemic risk is hard to 
reduce with traditional regulation schemes \cite{thurner09,pol}. 
By introducing a minimum level of transparency in financial networks, endogenous risk can be drastically 
reduced without negative effects on the efficiency in the financial services for the real economy. 
 
 \begin{figure}
	\begin{center}
		\includegraphics[width=8cm]{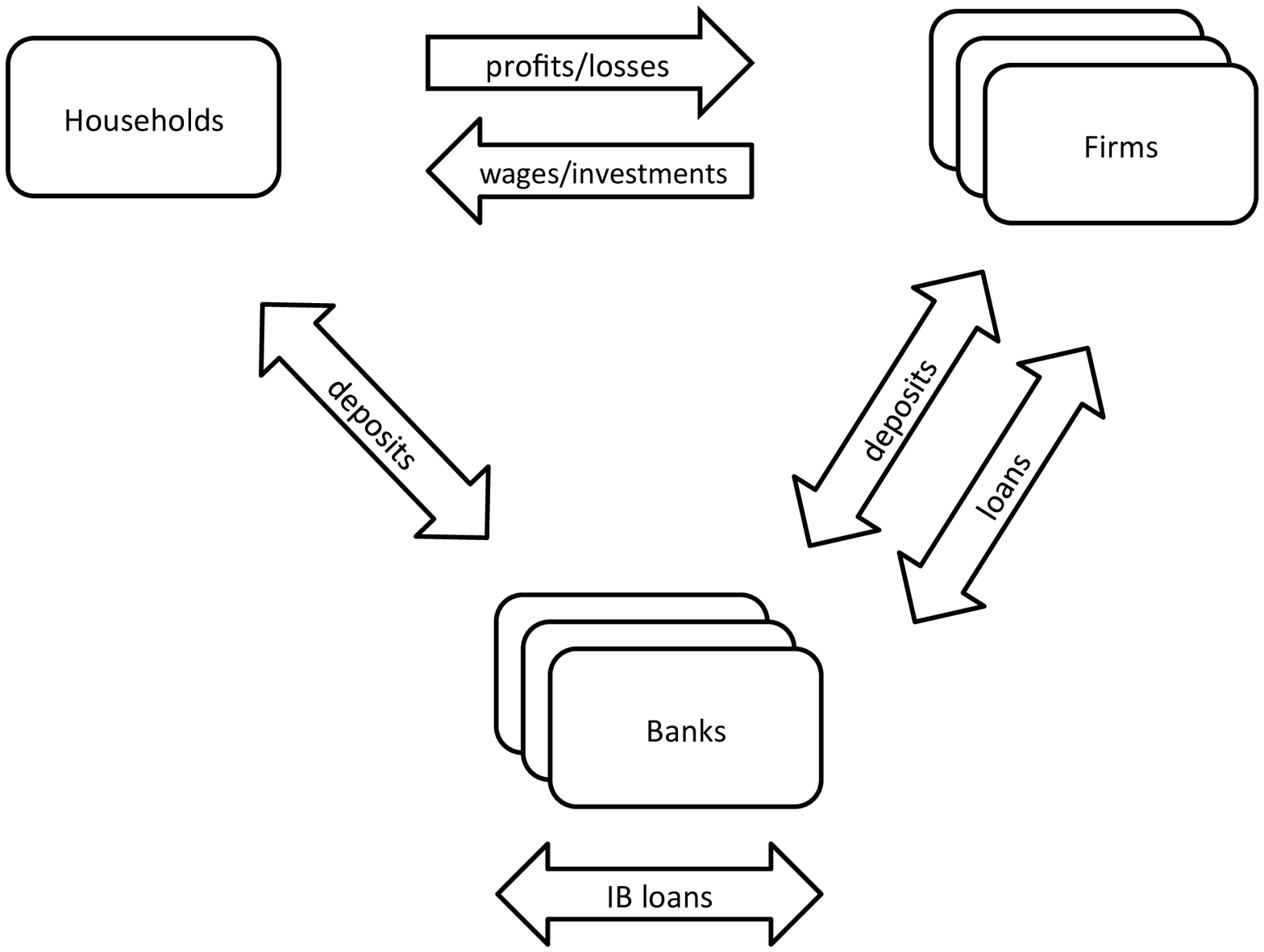} 
		\includegraphics[width=8cm]{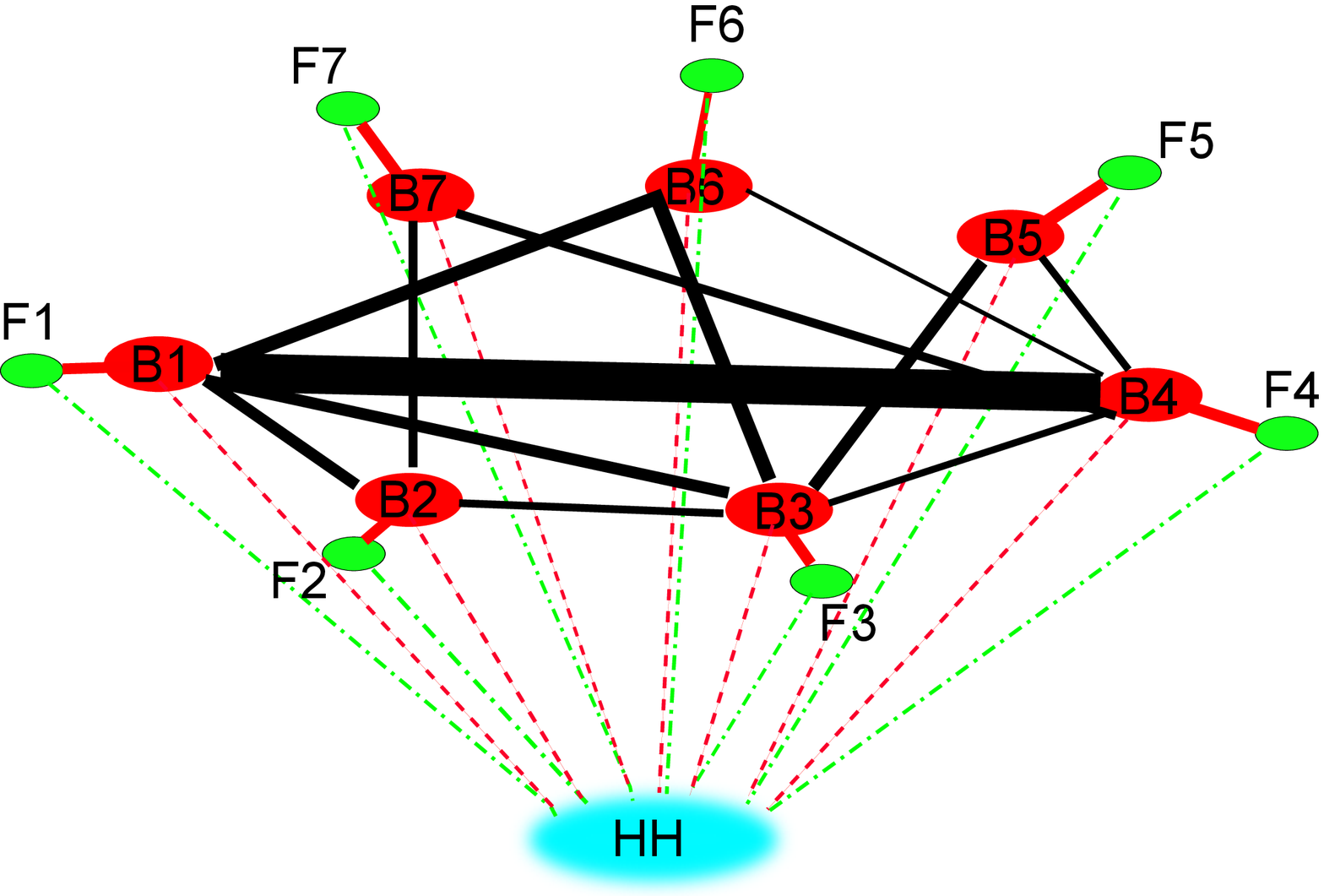} 
	\end{center}
	\caption[Fig1]{
	Schematic structure of the model: firms approach their bank for firm-loans. These loans are transferred to the households, 
	where they are redistributed to other firms or are deposited in banks. If a bank can not service a firm-loan because it 
	presently does not  have the requested sum available in the bank, it tries to get the necessary funding through an interbank (IB) 
	loan from an other bank. If it can get an IB loan, the firm-loan is payed out, if it can not,  the firm will receive no loan. 
	} \label{fig1} 
\end{figure}

In most developed countries interbank  loans are recorded in the `central credit register' of Central Banks, that reflects the asset-liability 
network of a country \cite{freparroc00}. The capital structure of  banks is available through standard reporting to 
Central Banks. Payment systems record financial flows with a time resolution of one second, see e.g. \cite{target}. 
Several studies have been carried out on historical data of  asset-liability networks \cite{boss,uppwor02,LNCS2004,brasil,fed,fin},  including 
overnight markets  \cite{iori}, and financial flows \cite{fragyskos}.

Given this data, it is possible (for Central Banks) to compute network metrics of the asset-liability matrix in real-time, 
which in combination with the capital structure of banks, allows to define a systemic risk-rating of banks. 
A systemically risky bank in the following is  a bank that -- should it default -- will have a substantial impact 
(losses due to failed credits) on other nodes in the network.  
The idea of network metrics is to systematically capture the fact, that by borrowing from a systemically risky bank, the borrower 
also becomes systemically more risky since its default might tip the lender into default. 
These metrics are inspired by PageRank, where a web page, that is linked to a famous page, gets a share of the `fame'. 
A metric similar to PageRank,  the so-called DebtRank, has been recently used to capture systemic risk levels in 
financial networks \cite{battiston12}. 
In this paper we present an agent based model of the interbank network that allows to estimate the extent to which 
systemic risk can be reduced by introducing transparency on the level of the DebtRank. For computational efficiency we propose a 
measure based on Katz centrality \cite{katz}, which we refer to Katz rank. 
Both are closely related to the concept of eigenvalue centrality \cite{newman_book}. 
Betweenness centrality has been used to determine systemic financial risk before \cite{LNCS2004}. 
To demonstrate the risk-reduction potential of feeding information of the  DebtRank back into the system, 
we use a simple  toy model of the financial- and real economy which is described in the next section. 
Interbank models of similar design were used before in different contexts \cite{thurner2003,iori_simulation}. 

The central idea of this paper is to operate the financial network in two modes. The first reflects the situation today, 
where banks don't know about the systemic impact of other banks, and where all interbank (IB) credits 
are traded with the same interest rate, the so-called `inter bank offer rate', $r^{\rm ib}$. 
We call this scenario the {\em normal mode}.

The second mode introduces a minimum regulation scheme, where banks chose their IB trading partners 
based on their DebtRank. The philosophy of this scheme comes from the fact that borrowing from a systemically dangerous 
node can make the borrower also dangerous, since she inherits part of the risk, and thereby increases overall systemic risk. 
Note, that a default of the borrower from a systemically dangerous bank affects not only the lender, but possibly also 
all other nodes from which the lender has borrowed. The idea is to reduce systemic risk in the IB network by not 
allowing borrowers to borrow from risky nodes. 
In this way systemically risky nodes are punished, and an incentive for nodes is established to be low in systemic riskiness. Note, 
that lending {\em to} a systemically dangerous node does {\em not} increase the systemic riskiness of the lender.  
We implement this scheme by making the DebtRank of all banks visible to those banks that want to borrow. 
The borrower sees the DebtRank of all its potential lenders, and is required (that is the regulation part) 
to ask the lenders for IB loans in the order of their inverse DebtRank. 
In other words,  it has to ask the least risky bank first, then the second risky one, etc. In this way the most risky 
banks are refrained from (profitable) lending opportunities, until they reduce their liabilities over time, which makes them  
less risky. Only then will they find lending possibilities again. 
This mechanism has the effect of distributing risk homogeneously through the network, and prevents the emergence 
of systemically risky nodes in a self-organized critical way: risky nodes reduce their credit risk because they are 
blocked from lending, non-risky banks can become more risky by lending more. 
We call this mode the {\em transparent mode}. 
 
\section{The model} \label{model}

The agents in the model are $B$ banks, $F=B$ firms and households. For details of the implementation, see SI\footnote{http://www.complex-systems.meduniwien.ac.at/people\\/sthurner/SI/DebtRank/katz\_rank\_SI\_05.pdf}. 
At every timestep each firm approaches its main bank with a request for a loan. The size of these loans is a random number  
from a uniform distribution. 
Banks try to provide the requested firm-loan. If they have enough cash reserves available, the loan is granted. If they do not have enough, 
they approach other banks in the interbank (IB) market and try to get the amount from them at an interest rate of $r^{\rm ib}$. 
Not every bank has business relations with every other bank.
Interbank relations are recorded in the IB {\em relation network}, $A$. If two  banks 
$i$ and $j$, are wiling to borrow from each other, $A_{ij}=1$, if they have no business relations, $A_{ij}=0$. We model the 
IB relation network with random graphs and scale-free networks, see SI. If a bank does not have enough cash and can not raise the 
full amount for the requested firm-loan on the IB market, it does not pay out the loan. 
If the bank does pay out a loan, the firm transfers some of the cash to the households as `investments' for future payoffs  
(wages,  invest in new machines, etc.). 
Loans from previous timesteps are paid back after $\tau$ timesteps with an interest rate of $r^{\rm f-loan}>r^{\rm ib}$. 
The fraction of the loan not used to pay back outstanding loans, ends up at the households (for details see SI). 

Households use the money received from firms 
to (1) deposit a certain fraction  at the bank, for which they get interest of $r^{\rm h}$, 
or (2) to consume goods produced by other firms (details in SI). This money flows back to firms (the firms' profits) 
and is used by those to repay  loans. If firms run a surplus, they deposit it in their bank accounts, receiving interest of $r^{\rm f-deposit}$. 
The two actions of the households effectively lead to a re-distribution and re-allocation of funds at every timestep. 
For simplicity we model the households as a single (aggregated) agent that receives cash from firms (through firm-loans) and 
re-distributes it randomly in banks (household deposits), and among other firms (consumption). 

Specifically, at time $t$ a bank-firm pair is chosen randomly, and the following actions take place:
\begin{itemize}
\item[(i)] banks and firms repay loans issued at time  $t-\tau$ 
\item[(ii)] firms realize profits or losses (consumption)
\item[(iii)] banks pay interest to households
\item[(iv)] firms request loans 
\item[(v)]  households re-distribute cash obtained from firms 
\item[(vi)]  liquidity management of banks in the IB market, including:
	 IB re-payments,  firm-loan requests,  defaulted firms, and re-distribution effects from households
\item[(vii)] firms pay salaries and make investments
\item[(viii)] firms or banks default if equity- or liquidity problems arise
\end{itemize}
A new bank-firm pair is picked until all are updated (random sequential update); then timestep $t+1$ follows.

During the simulation, firms and banks may be unable to pay their debts and thus become insolvent. 
Firms are declared bankrupt if they are either insolvent, or if their equity capital falls below some negative threshold. 
Banks are declared bankrupt if they are insolvent, or have equity capital below zero. 
If a firm goes bankrupt the bank writes off the respective outstanding loans as defaulted credits and realizes the {\em losses}. 
If the bank has not enough equity capital to sustain these losses it goes bankrupt as well. 
After the bankruptcy of a bank there occurs a default-event for all its IB creditors.
This may trigger a cascade of bank defaults. For simplicity, there is no recovery for IB loans. 
A cascade of bankruptcies happens within one timestep. After the last bankruptcy is taken care of, the simulation is stopped.
We model a {\em closed} system of banks, firms and households, meaning that there are no in- or out-flows of cash from the model.

\subsection{The normal mode}

In the normal mode the model captures the current market practice, where  
banks follow a simple strategy to manage their liquidity. If a bank needs additional liquidity (for 
providing a firm-loan request, or for its own re-payments of IB loans) it contacts banks 
it is connected with in the IB relation network $A$, and asks them for IB loans. In the 
normal mode, bank $i$ asks its neighbors $j$ (with $j \in {\cal I}_{i} = \{j' \mid A_{ij'}=1 \}$) in {\em random} order.
If bank $j$ can provide only a fraction of the requested IB loan, bank $i$ takes it, and continues to ask 
an other neighbor bank from ${\cal I}_{i}$ (in random order) until the liquidity requirements of $i$ are satisfied. 

\begin{figure*}
	\begin{center}
		\includegraphics[width=5.9cm]{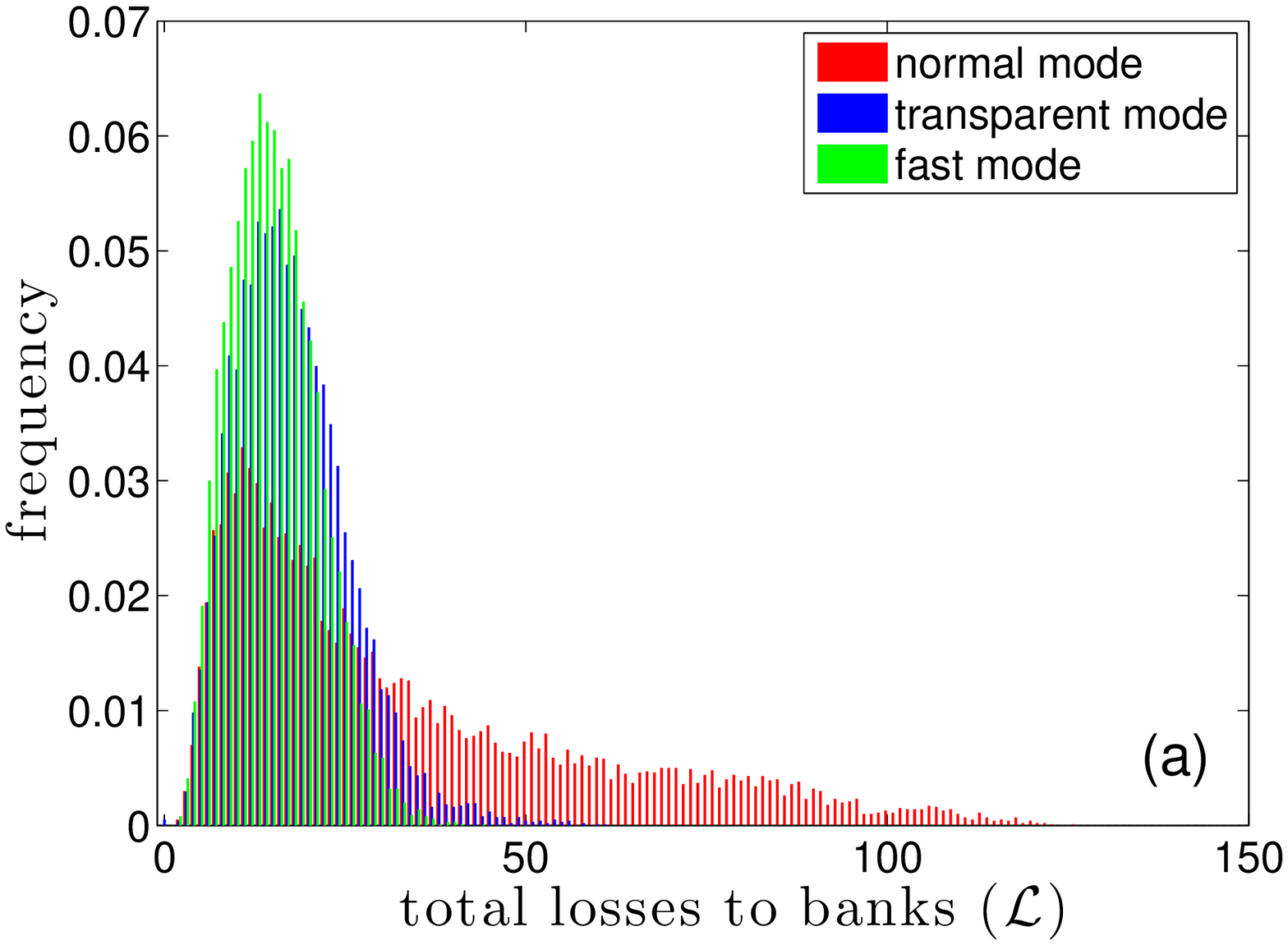} 
		\includegraphics[width=5.9cm]{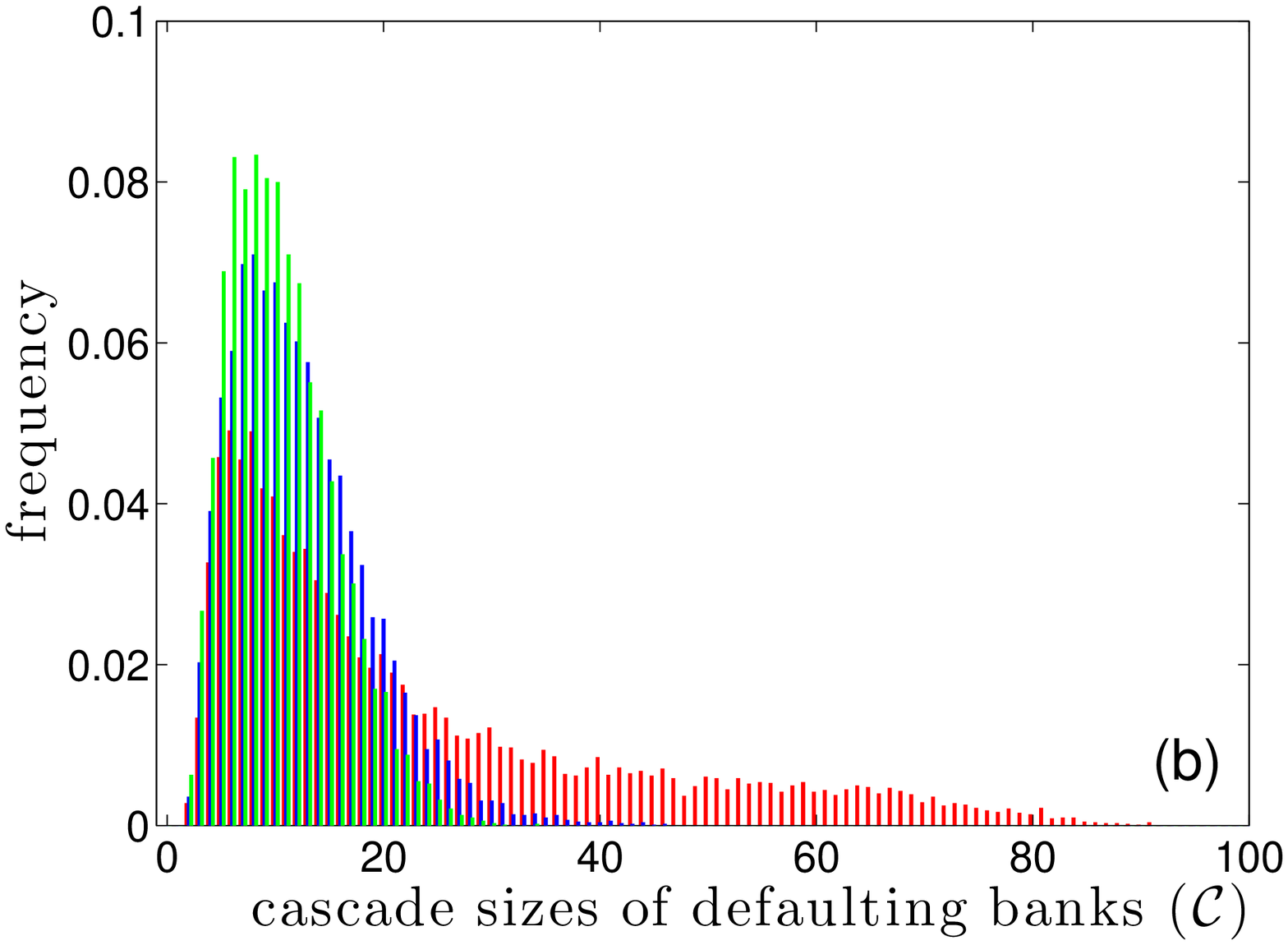} 
		\includegraphics[width=5.9cm]{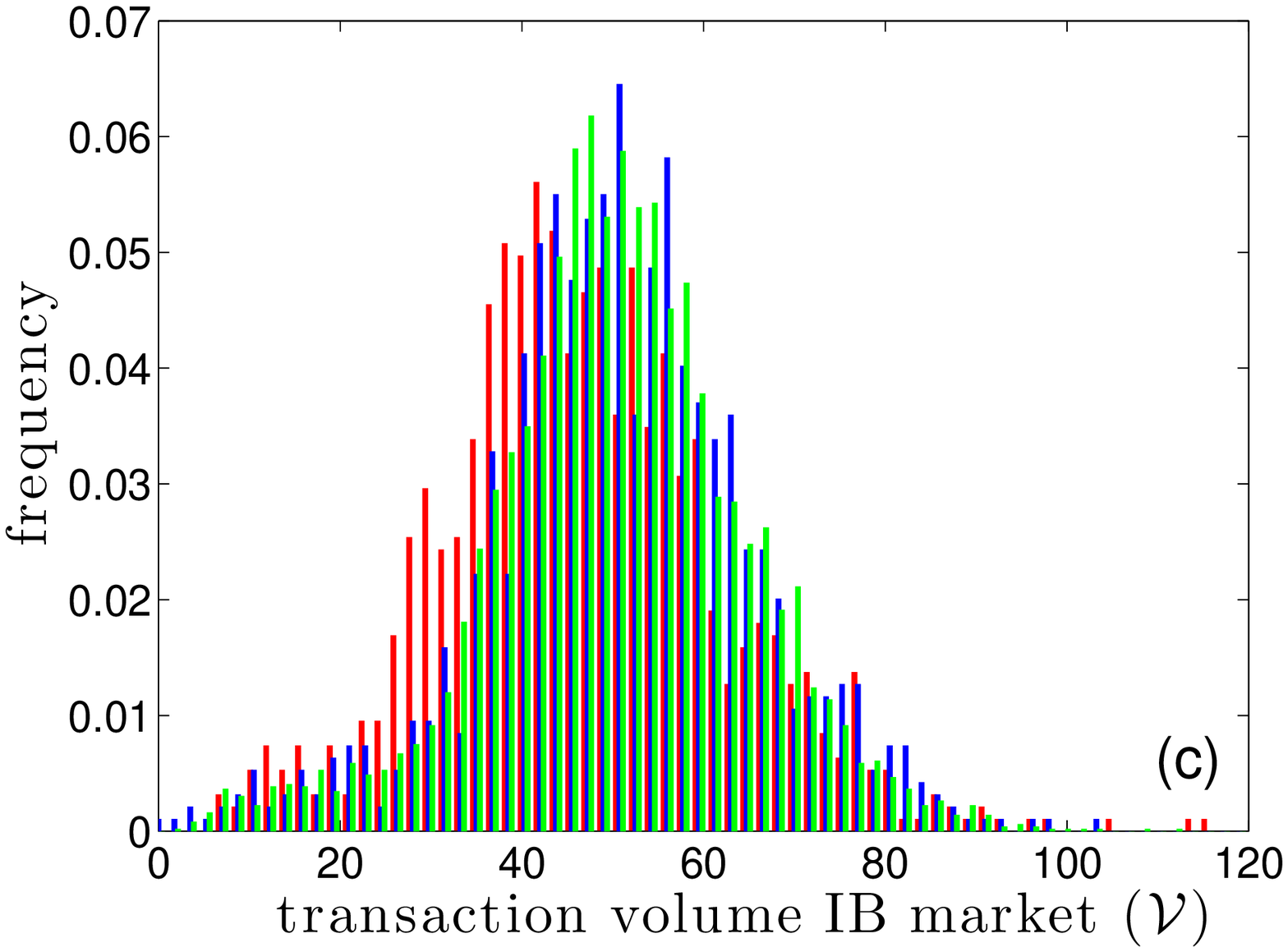} 
	\end{center}
	\caption[] {Comparison of the normal mode (red), i.e. random selection of counterparty for IB loans, 
		 with  the transparent IB market (blue), where the order of counterparty selection is determined by the inverse DebtRank.
		 The fast mode is shown in green.
	(a) Distribution of total losses to banks ${\cal L}$,  
	(b) distribution of cascade sizes ${\cal C}$ of defaulting banks, and 
	(c) distribution of transaction volume in the IB market ${\cal V}$. 
	 We performed 10,000 independent, identical  simulations, each with 500 timesteps, 100 banks, and the simulation parameters given in SI Table I. 
	 $A$ is a fully connected graph. 
	 }
	 	\label{hist_loss} 
\end{figure*}

\subsection{The transparent IB mode} 

A simple modification to improve the stability of the system is to avoid borrowing from banks with a large systemic 
impact. For this a minimum level of transparency of the  IB market is necessary. 
For all banks we compute systemic risk metrics  based on the interbank liability network $L_{ij}(t)$, and the 
equity of banks $C^{\rm b}_{i}(t)$, at timestep $t$ (details in SI). 
In particular we compute the DebtRank $R^{\rm debt}_{i}$ \cite{battiston12}, and -- for comparison -- the 
Katz rank $R^{\rm katz}_{i}$ (see methods). The most risky bank has rank 1, the least risky has rank $B$, 
see methods.

In contrast to the normal mode,  before bank $i$ asks its neighbors 
for IB loans, it orders them (the banks contained in set  ${\cal I}_{\rm i}$) according to their inverse Debt- or Katz rank.   
It then asks its neighboring banks in the order of their inverse rank, i.e. it first asks the least risky, then the 
next risky, etc. The rank is computed at the beginning of each timestep. 
In this way the low-risk banks are favored because the likelihood for obtaining (profitable) IB deals is much 
higher for them than for risky banks, which are at the end of the list and will practically never be asked. 

In reality this implies that the banks  know the DebtRank of each of their neighboring banks. This transparency 
is not available in the present banking system.  Note however, that in many countries Central Banks have all the 
necessary data to compute the DebtRank. A possible way to implement such an incentive scheme in reality, 
is presented in the discussion. 

\subsection{Transparent IB market -- fast mode} 
We implement a version of the transparent IB market, where the DebtRank is computed after 
every transaction that takes place in the IB market, instead of being computed at the beginning of the day. 
This version we refer to as the {\em fast mode}.

\section{Results} \label{results}

We simulate the above model with the parameters given in SI, for 500 timesteps. Results are averages over 10,000 
identical simulations. Fit parameters to the following distribution functions are collected in SI Table 2. 

In Fig. \ref{hist_loss} (a) we show the distribution of losses to banks ${\cal L}$ for the 
the normal mode (red), where the selection of counterparties for IB loans is  random and the transparent mode (blue), 
where banks sort their potential counterparties according to their inverse DebtRank, and then approach the least risky 
neighbor first for the IB loan. The fast mode is shown in green. 
The normal mode shows a heavy tail in the loss distribution, which completely disappears 
in the  transparent and fast modes, where there are no losses higher than 50 and 40 units, respectively. 
Of course losses do not entirely disappear in the transparent scheme, since the credit risk that firms bring to the banking 
system can not be completely eliminated. 
The fast mode appears to be slightly safer than the transparent mode. Fits to all curves are found in SI. 

The distribution of cascade-sizes  ${\cal C}$ of defaulting banks is seen in Fig. \ref{hist_loss} (b). Again the normal mode shows a 
heavy  tail, meaning that in a non-negligible number of events, defaults of a single bank trigger a cascade of liquidity 
and equity problems through the system. In some cases up to 80 \% of the banks collapse.     
In the transparent mode the likelihood for contagion is greatly reduced, and the maximum cascade size is limited by 40 
banks in the transparent and about 30 in the fast mode.   

In Fig. \ref{hist_loss} (c) we show the transaction volume in the IB market ${\cal V}$ of the three modes, normal  (red), 
transparent  (blue) and fast (green). The transparent and fast modes  show a higher transaction volume indicating 
a more efficient IB market, where liquidity from banks with excess funds is more effectively channeled to those without. 
We verified, that the ratio of requested- to provided firm-loans, the efficiency ${\cal E}$,  yields ${\cal E}\sim 1$, 
irrespective of the mode. 

\begin{figure}
	\begin{center}
		\includegraphics[width=7cm]{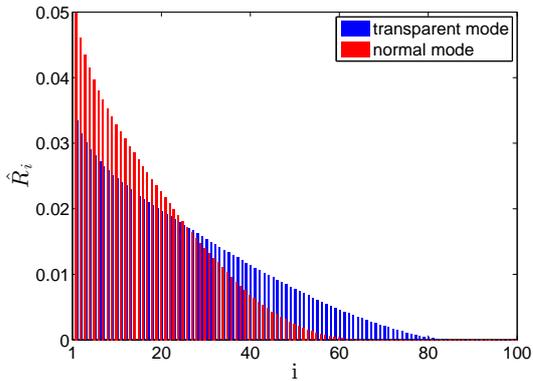} 
	\end{center}
	\caption[] {
	Normalized DebtRank, $\hat{R_{\rm i}}$ for individual banks in the normal (red) and the transparent mode (blue).
	Banks are ordered according to their DebtRank, the most risky  is to the very left, the safest  to the very right. 
	The distribution is an average over $1000$  simulation runs with an ER network, and shows the situation at timestep $t=100$. 
	} 
\label{katzdistribution} 
\end{figure}

In  Fig. \ref{katzdistribution} we show the normalized DebtRank for all individual banks, for the normal (red), and the transparent scheme (blue). 
Banks are rank ordered according to their DebtRank so that the most risky bank is found to the very left, the safest to the very right. 
It is clear that the systemic risk impact in the transparent mode is spread more  evenly throughout the system, 
whereas in the normal mode some banks appear to be much more dangerous to the system.

\begin{figure}
	\begin{center}
		\includegraphics[width=7.0cm]{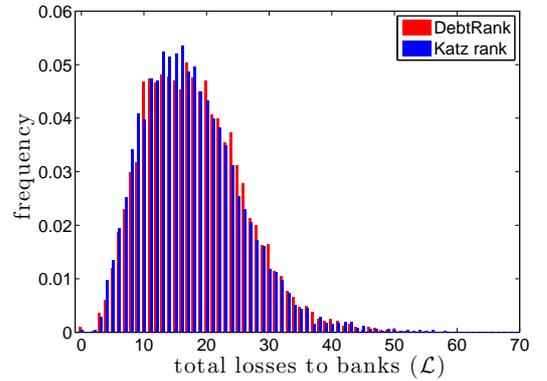} 
	\end{center}
	\caption[] {
	Comparison of the performance of the DebtRank (red) and the  Katz rank (blue) for the losses, ${\cal L}$.  
	Both rank definitions provide practically identical results. 	
	Same simulation parameters as in previous figure.
	 } 
\label{katz_debt} 
\end{figure}

In Fig. \ref{katz_debt}  we compare the losses ${\cal L}$ for DebtRank (red) and  Katz rank (blue). 
The performance of of the two definitions is hardly distinguishable. 
Also  the other systemic risk measures show no noticeable difference, for cascade size ${\cal C}$, and 
transaction volume distributions ${\cal V}$, see SI.  

\begin{figure}
	\begin{center}
		\includegraphics[width=7cm]{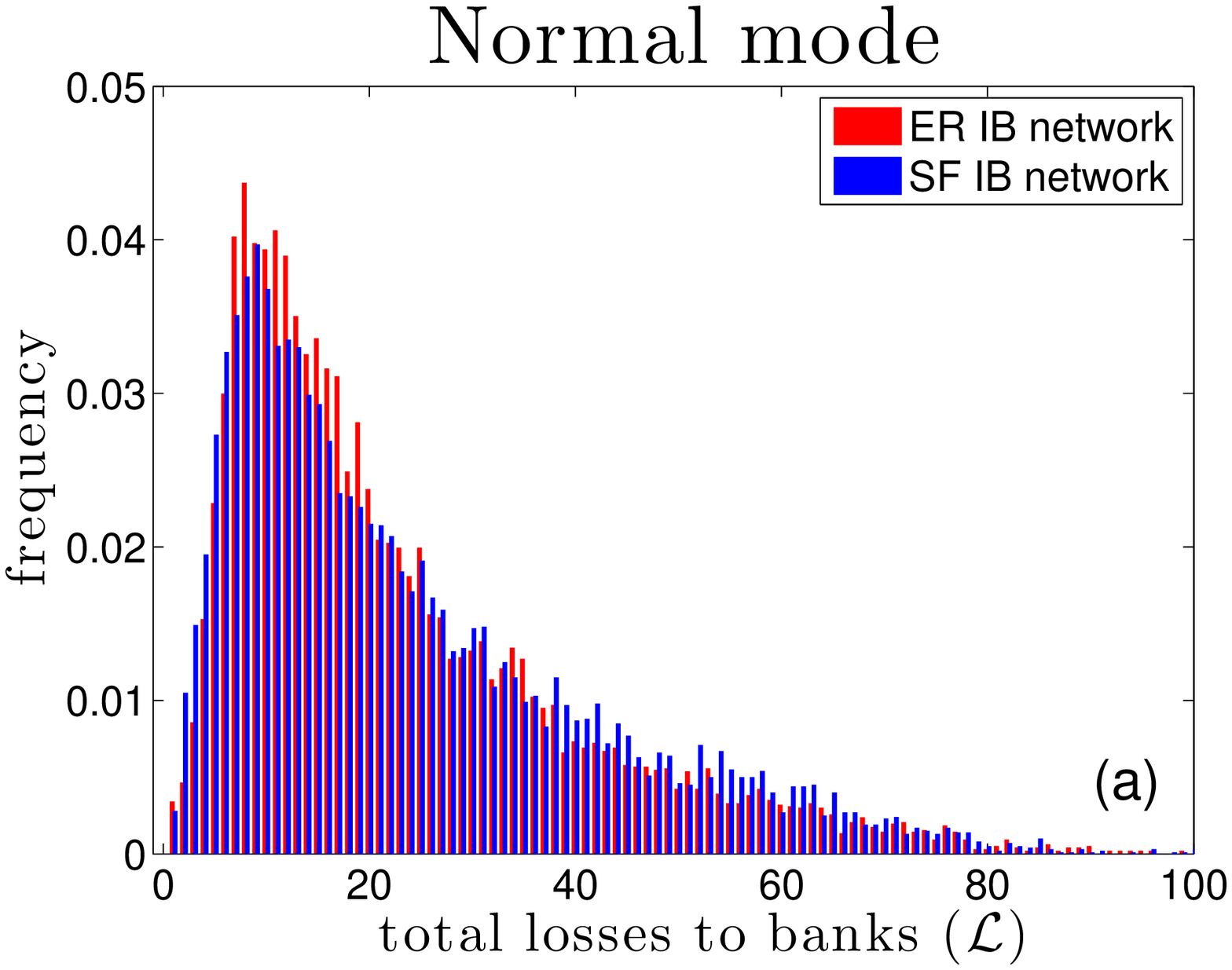} 
		\includegraphics[width=7cm]{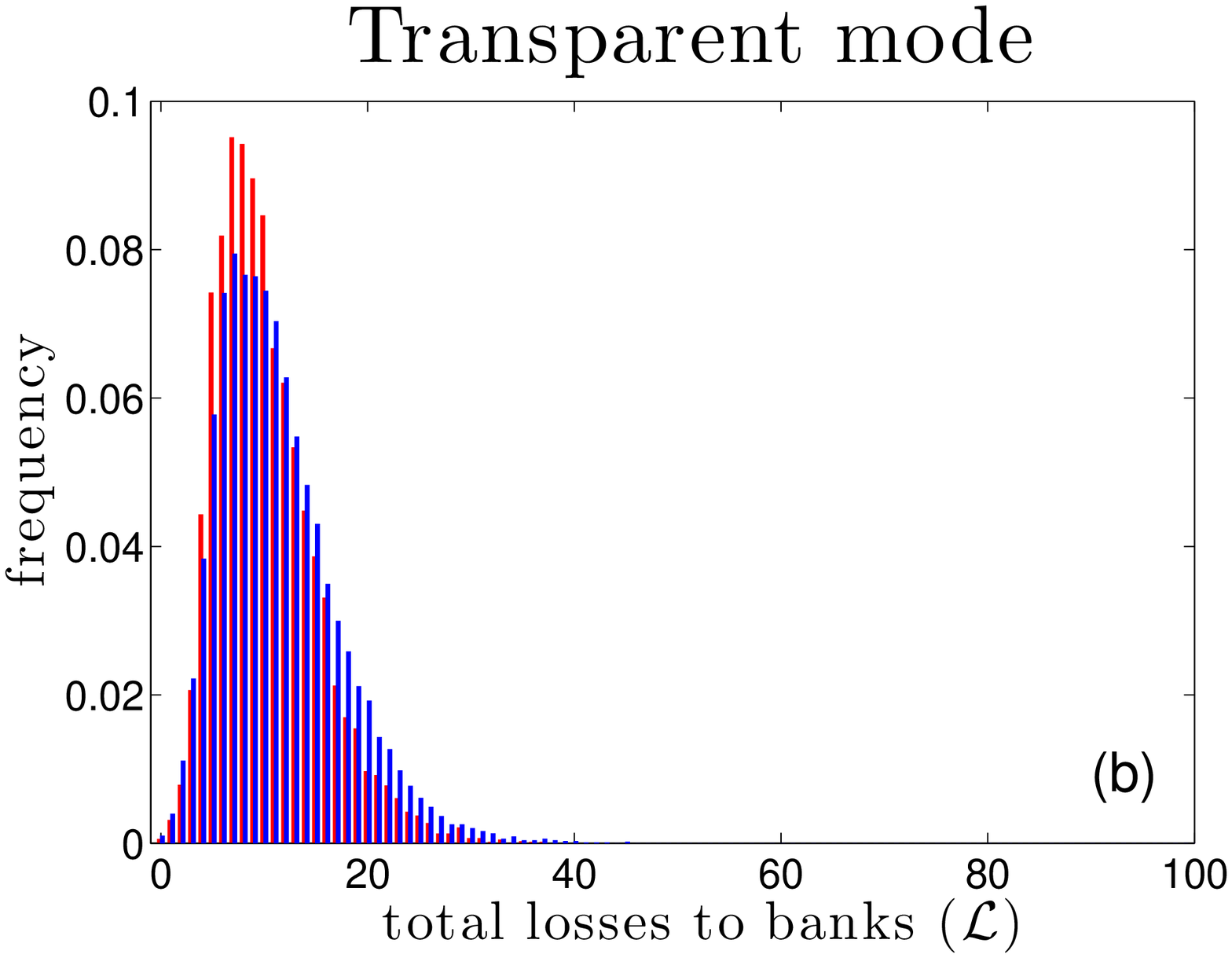} \\
	\end{center}
	\caption[]{Distributions of  losses ${\cal L}$ for the normal (a) and transparent mode (b),  
	for an ER  (red) and a  SF network (blue), both with the same average connectivity  $\langle k \rangle = 11.5$. 
	} 
\label{NW_effects} 
\end{figure}

Figure \ref{NW_effects}  shows the distribution of  losses  ${\cal L}$, for the (a) normal 
and  (b) transparent mode, 
as computed with an ER contact network (red) with $\gamma=0.115$, and a scale-free BA network (see SI) with the same average 
connectivity ($\langle k \rangle = 11.5$). 
In both modes the SF network leads to a slightly riskier situation. 
The situation for cascade sizes and transaction volume is depicted in SI Fig. 2, where we also show and discuss 
the effects of connectivity on the three measures in SI Fig. 3.

Finally, we compute the distribution of the time to first default ${\cal T_{\rm fd}}$, for the normal and the transparent modes. 
Both distributions are practically Gaussian (kurtosis $\sim 3.3$, skewness $\sim 0.4$) with mean and standard deviation of 
${\cal T}_{\rm fd}^{\rm normal}=138.2 \pm 33.8$, 
and ${\cal T}_{\rm fd}^{\rm transp.}=138.1 \pm 33.7 $, respectively.  
This is expected, since typically the first default is triggered by a firm-default, which is (to first order) independent of the situation in the 
IB market, but only depends on the parameters describing the firms ($\mu_{ i}$, $\sigma^{\rm return}$, $C^{\rm default}$) 
and households ($\sigma$, $\rho$), see SI. 

\section{Methods} \label{methods}

\subsection{DebtRank} \label{debtrank}
DebtRank is a recursive method suggested in \cite{battiston12} to determine the systemic relevance of nodes in financial networks. 
It is a number measuring the fraction of the total economic value in the network that is potentially affected by a node or a set of nodes. 
Let $L_{ij}$ denote the IB liability network at a given moment (loans of bank $j$ to bank $i$), and $C_{i}$ 
is the capital of bank $i$, see SI. If bank $i$ defaults  and can not repay its loans, bank $j$ 
loses the loans $L_{ij}$, see SI. If $j$ has not enough capital available to cover the loss, $j$ also defaults. 
The impact of bank $i$ on bank $j$ (in case of a default of $i$) is therefore defined as 
\begin{equation}
	\label{impact} W_{ ij} = \min \left[1,\frac{L_{ ij}}{C_{ j}} \right] .
\end{equation}
Given the total outstanding loans of bank $i$, $L_{ i}=\sum_{ j}L_{ ji}$, 
 its {\em economic value} is defined as $v_{i}=L_{ i}/\sum_{ j}L_{ j}$. 
The value of the impact of bank $i$ on its neighbors is $I_{i} = \sum_{ j} W_{ ij} v_{ j}$. 
To take into account the impact of nodes at distance two and higher, it has to be computed recursively, 
\begin{equation}
	\label{recursive_impact} I_{i} = \sum_{\rm j} W_{\rm ij} v_{\rm j} + \beta \sum_{\rm j} W_{\rm ij} I_{\rm j} ,
\end{equation}
where $\beta$ is a damping factor.
If the network $W_{ ij}$ contains cycles the impact can exceed one. 
To avoid this problem an alternative was suggested \cite{battiston12}, where  two state variables, $h_{\rm i}(t)$ and $s_{\rm i}(t)$, are assigned 
to each node. $h_{\rm i}$ is a continuous variable between zero and one; 
$s_{\rm i}$ is a discrete state variable for 3 possible states, undistressed, distressed, and inactive, $s_{\rm i} \in \{U, D, I\}$.
The initial conditions are $h_{ i}(1) = \Psi \, , \forall i \in S_f ;\; h_{ i}(1)=0 \, , \forall i \not \in S_f$, and 
$s_{ i}(1) = D \, , \forall i \in S_f ;\; s_{ i}(1) = U \, , \forall i \not \in S_f$ (parameter $\Psi$ quantifies  
the initial level of distress: $\Psi \in [0, 1]$, with $\Psi = 1$ meaning default).  The dynamics of $h_i$ is  
then specified by
\begin{equation}
	h_{ i}(t) = \min\left[1,h_{ i}(t-1)+\sum_{j\mid s_{ j}(t-1) = D}  W_{ ji}h_{ j}(t-1) \right] . 
	\end{equation}
The sum extends over these $j$, for which $s_{ j}(t-1) = D$, 
\begin{equation}
	s_{ i}(t) = 
	\begin{cases}
		D & \text{if } h_{ i}(t) > 0; s_{ i}(t-1) \neq I ,\\
		I & \text{if } s_{ i}(t-1) = D , \\
		s_{ i}(t-1) & \text{otherwise} .
	\end{cases}
\end{equation}
The DebtRank of set $S_f$ (set of nodes in distress at time $1$), is  
$R = \sum_{ j} h_{ j}(T)v_{ j} - \sum_{ j} h_{ j}(1)v_{ j}$,  
and measures the distress in the system, excluding the initial distress. 
If $S_f$ is a single node, the DebtRank measures its systemic impact on the network. 
The DebtRank of $S_f$ containing only the single node $i$ is 
\begin{equation}
	\label{debtrank} R_{ i} = \sum_{ j} h_{ j}(T)v_{ j} - h_{ i}(1)v_{ i} . 
\end{equation}
The normalized DebtRank is $\hat R_i = R_i/\sum_i R_i$.

\subsection{Katz rank} \label{katz}
The Katz centrality can be used to  capture the risk of contagion in IB networks and is defined as
\begin{equation}
	\label{katz} K_{i} = \alpha \sum_{ j} L_{ ij} K_{ j} + \beta, 
\end{equation}
To ensure that the Katz centrality converges, we set $\alpha = 1/\kappa_1$, where $\kappa_1$ is the largest eigenvalue of $L_{ij}$. 
The overall multiplier $\beta$ is not important, for convenience we set $\beta = 1$.
We define the {\em Katz rank} in the following way:
the most risky bank $i$, with the highest Katz centrality gets Katz rank $R^{\rm katz}_{ i}(t)=1$, 
the least risky bank $j$ (lowest Katz centrality) gets  $R^{\rm katz}_{ j}(t)=B$.
Note, that banks that only borrow (in-links only) may cause contagion and have non-zero Katz centrality. 
Banks that only provide loans (out-links) have zero Katz centrality.  
Loan sizes, neighbors and their neighbors, etc., are included in the centrality.

\subsection{Systemic Risk measures} \label{risk_measures}
For  measures of systemic risk in the system, we use  
the following three observables:
(1) {\em time to first default} as the timesteps of the simulation before the first default of a bank occurs, ${\cal T}^{\rm fd}$, 
(2) the {\em size of the cascade}, ${\cal C}$ as the number of defaulting banks triggered by an initial bank default ($1\leq {\cal C} \leq B$), 
and (3) the {\em total losses to banks} following a default or cascade at $t_0$, ${\cal L}= -\sum_{i}^B [C_{\rm i}(t_0)-C_{\rm i}(t_0-1)]$. 

\subsection{Efficiency measures} \label{efficiency_measures}

To quantify  the efficiency of the banking system we use the ratio of the  sum of requested loans by the firms to the 
sum of loans actually payed out to firms, at a given time $t$, averaged over time, 
 \begin{equation}
 	E(t)=  \frac{\sum_{i=1}^B  l_{ i}(t)}{\sum_{i=1}^B l_{ i}^{\rm req}(t)}  .
 \end{equation} 
The efficiency of the system is then the time average ${\cal E}=  \langle  E(t) \rangle_t$, taken over the simulation.
As another measure of efficiency we use the transaction volume in the IB market at a particular time $t=T$ in a typical simulation run, 
 \begin{equation}
	\mathcal{V}(t)=\sum_{j=1}^B\sum_{i=1}^Bl_{ ji}(T)+l_{ ji}(T-\tau) .
 \end{equation}
The first term represents the new IB loans at timestep $T$, the second the loans that are repaid. We set $T=100$.

\section{Discussion} \label{discussion}
We showed that the risk, endogenously created in a financial network by the inability of banks to carry 
out  correct risk-estimates of their counterparties, can be drastically reduced by introducing a minimum level of transparency. 
Systemic risk is significantly reduced by introducing an incentive that makes borrowers more prone to borrow 
from systemically safe lenders. This philosophy was implemented by 
making a centrality measure such as the DebtRank available to all nodes in the network at each point in time.   
We could show that the efficiency of the financial network with respect to the real economy is 
not affected by the proposed regulation mechanism (${\cal E} \sim 1$ in all modes). 
This is possible since the regulation only re-distributes risk in order to avoid 
the emergence of risky agents that might threaten the system, and does not reduce the trading volume on the IB network. 
Risky nodes that are low in DebtRank, are barred from the possibility of lending their excess reserves to others. 
This deprives them from making profits on IB loans, but also reduces their risk of being hit by defaulted credits.  
They only receive payments and do not issue more risk, meaning that over time they become less risky. Less risky banks are allowed to 
take more risk (lend more) and make more profits. The proposed mechanism makes the system safer in a self-organized critical manner.   

Note, that in our scheme the borrower determines who borrows from whom. 
Usually the lender is concerned if the borrower will be able to repay the loan. However, this risk of credit default is not necessarily of systemic relevance.
Lending to a bank with a large systemic risk can have relatively little consequences for the 
systemic importance for the lender, or the systemic risk of the system as a whole. In contrast, if a bank borrows from a systemically 
dangerous node the borrower inherits part of this risk, and increases the overall systemic risk. 
These facts are conveniently incorporated in the definition of Debt- and Katz rank. 

We found that the performance of the method is surprisingly insensitive to the choice of the particular centrality measure, or whether 
the actual topology  of the IB network (scale-free or random).  Also the average connectivity $\bar k$ of the network is not relevant, as long 
as it remains in sensible regions ($ \langle k \rangle \in [\sim5,B]$). This suggests that the essence of the proposed scheme is that risk is spread 
more evenly across the network, which practically eliminates cascading failures.  

A possible way to  implement the proposed transparency in reality would be that Central Banks regularly compute the DebtRank 
and make it available to all banks. To enforce the regulation, the CB monitors the IB loans in their  central credit register, 
and severely punishes borrowers who failed to find less risky lenders.  To design a more market based mechanism to obtain the 
same self-organized critical regulation dynamics is subject to further investigations. 

{\bf Acknowledgements}
We thank P. Klimek for  stimulating conversations and a careful reading of the manuscript. 
Funding was provided by EU FP7 project CRISIS, agreement no. 288501. 


\end{document}